\documentclass[aps,pre,preprint,groupedaddress,amsmath,amssymb,showpacs]{revtex4-1}

\usepackage{graphicx}

\newcommand{\I}{\textup{i}}
\newcommand{\E}{\textup{e}}
\newcommand{\D}{\textup{d}}

 \newcommand{\ketbra}[2]{\mathopen{|}#1\mathclose{\rangle}\mathopen{\langle}#2\mathclose{|}}
 \newcommand{\op}[1]{\hat{#1}}

\newcommand{\dB}{\delta_B} 
\newcommand{\dS}{\delta} 
\newcommand{\Dd}{\Omega} 

\newcommand{\Ptot}[1]{\op{\tilde{P}}_{#1}} 
\renewcommand{\P}[1]{\op{P}_{#1}} 
\newcommand{\rtot}{\op{\rho}_\text{tot}} 
\newcommand{\rS}{\op{\rho}} 
\newcommand{\rB}{\op{\rho}_B} 
\newcommand{\rtotbef}{\op{\rho}_\text{tot}^\prime} 
\newcommand{\rSbef}{\op{\rho}^\prime} 
\newcommand{\rBbef}{\op{\rho}_B^\prime} 
\newcommand{\secordcomm}{\op{\mathfrak{c}}} 
\newcommand{\rc}[1]{\rho_{#1}} 
\newcommand{\avrc}[1]{\overline{\rho}_{#1}} 
\newcommand{\attr}[1]{\overline{\rho}_{#1}^\text{attr}} 
\newcommand{\ReOff}{\mathfrak{R}} 
\newcommand{\ImOff}{\mathfrak{I}} 
\renewcommand{\Re}[1]{\text{Re}\left(#1\right)}
\renewcommand{\Im}[1]{\text{Im}\left(#1\right)}
\newcommand{\sinA}{\zeta_1} 
\newcommand{\sinB}{\zeta_2} 
\newcommand{\Tr}[2][]{\text{Tr}_{#1}\!\left\{#2\right\}} 
\newcommand{\Dt}{{\Delta t}} 
\newcommand{\eu}{\,u}
\newcommand{\gam}{\gamma}

\newcommand{\kom}[2]{\left[#1,#2\right]}
\newcommand{\dnd}[2]{\frac{\D #1}{\D #2}} 

\begin{document}


\title{Effective environments: Preparation of stationary states with inverse temperature ranging from positive to negative values}



\author{T. Jahnke}
\email[]{thomas.jahnke@itp1.uni-stuttgart.de}
\author{G. Mahler}
\affiliation{Institut f\"{u}r Theoretische Physik 1, Universit\"{a}t
             Stuttgart, Pfaffenwaldring 57, D-70550 Stuttgart, Germany}


\date{\today}

\begin{abstract}
In this paper, we discuss how effective environments incorporating periodic measurements can be used to prepare a two-level system (TLS) in almost arbitrary thermal states:
Concretely, we study a TLS coupled to a spin environment, the magnetization of which is measured periodically.
In ensemble average these measurements cause a relaxation of the TLS into a thermal (diagonal) state. By adjusting the time between the 
measurements and the detuning of the environmental spins, the creation of very low temperatures as well as inversion becomes possible.
Our analytical results derived for large environments are numerically shown to be valid even for quite small environments,
down to only a few spins.
\end{abstract}

 \pacs{05.70.Ln, 05.30.-d, 42.50.Dv}

\maketitle


\section{Introduction}
Within a quantum-thermodynamical approach~\cite{Gemmer2009} a single subsystem can be shown to relax to a thermal state if 
appropriately embedded in a quantum environment: The pertinent temperature is determined by the spectral density of this environment,
which, for sufficiently large modular embeddings and for a given working point in energy space, shows an approximate exponential
behavior. Beyond a design of this density there appears to be no immediate possibility of further manipulation.

Putting aside explicitly time-dependent Hamiltonians, an interesting way to circumvent this limitation consists in the application of quantum measurements.
Measurements play an outstanding role in quantum mechanics, since they provide a connection between the abstract theory and 
experimentally accessible quantities. However, in contrast to classical measurements,
quantum measurements usually influence the measured system. A pertinent 
example for the influence of periodic measurements is the well-known Zeno effect~\cite{Misra1977,Itano1990}, i.e., the suppression
of decay by fast repeated measurements. Recently, also
the possibility of cooling a TLS coupled to an oscillator bath by applying periodic quantum non demolition measurements
has been discussed~\cite{Erez2008,Gordon2009,Bensky2010,Gordon2010}.

In a recent Letter~\cite{Jahnke2010} we discussed the influence of periodic measurements in a quantum-thermodynamical
setting and pointed out how different concepts of statistical mechanics may arise from this ``observed quantum thermodynamics''.
Here, we present a detailed discussion of the effect of periodic measurements in a generalized environment model, allowing for a 
detuning between system and environment. As we will see, this permits us to prepare the embedded system in almost arbitrary thermal
states --- even the creation of inversion becomes possible.


\section{Model \label{sec:model}}
A typical quantum-thermodynamical model consists of the system of interest and its environment, with which it is allowed to exchange
energy via a weak coupling.
The model we investigate here belongs to the class of systems with a modular environment.
Following~\cite{Jahnke2010} we consider a system $S$ (a TLS with energy splitting $\dS$) described by the Pauli operator $\op{\sigma}_z$ and coupled via $\op{V}$ (strength $\lambda$) to a quantum environment $B$ with Hamiltonian $\op{H}_B$,
\begin{equation}
    \op{H}_\text{tot}=\frac{\delta}{2}\op{\sigma}_z\otimes \op{1}_B+\op{1}_S\otimes\op{H}_B+\lambda\op{V}\,.
\end{equation}

The environment consists of $n$ spins (e.g., a paramagnetic salt) with energy splitting $\dB\equiv\dS+\Dd$ each, i.e., we allow for
a detuning $\Dd$ between system and environment.
We will show below that this detuning $\Dd$ is key for preparing the TLS in almost arbitrary thermal states.

Without spin-spin interaction the degeneracy of the energy level $E_k=\dB k$ with $k$ spins up is given by $N_k={{n}\choose{k}}$.
Around a given $k_0$, the working point in energy space, the degeneracy structure can be approximated by an exponential of the form
\begin{equation}
\label{eq:degeneracy}
 N_k\approx\mathcal{N}e^{\beta E_k}\,.
\end{equation} 
For $n>k_0\gg1$ one finds (see Appendix \ref{sec:thermalizingenvironment})
\begin{equation}
\label{eq:beta}
 \beta(k_0)\approx\frac{1}{\dB}\ln\left( \frac{n}{k_0}-1\right)\,.
\end{equation} 
Including a weak interaction within the environment, each energy level broadens into an energy band of width $\Delta \varepsilon_k\ll\dB$.

For the interaction between system and environment we choose
\begin{equation}
\label{eq:V}
 \op{V}=\left( \op{\sigma}^{+}+\op{\sigma}^{-}\right) \otimes \op{B}\,,
\end{equation} 
where $\op{\sigma}^{+}=\ketbra{1}{0}$ is the creation operator and $\op{\sigma}^{-}=\ketbra{0}{1}$ the annihilation operator for the TLS
and the environmental part reads
\begin{eqnarray}
\label{eq:B}
 \op{B}=\sum_{k}\sum_{n_k,m_{k+1}}&&C_{k+1,k}(m_{k+1},n_{k})\ketbra{n_k}{m_{k+1}}\nonumber\\
&\quad+&C_{k,k+1}(n_k,m_{k+1})\ketbra{m_{k+1}}{n_k}\,.
\end{eqnarray} 
To keep the model as unbiased as possible, we do not further specify the interaction but consider the $C_{i,j}$ to be a set of 
random matrices whose entries are taken from a Gaussian distribution and are normalized to 
$\overline{|C_{i,j}(a,b)|^2}=(N_i N_j)^{-1/2}$. This choice of the normalization together with a sufficiently small $\lambda$
guarantees the weak coupling for arbitrary states.
Due to Hermiticity of $\op{V}$, we have $C_{i,j}=C_{j,i}^\dagger$.

Initially we assume a product state for the total system $\op{\rho}_\text{tot}(0)=\op{\rho}(0)\otimes\op{\rho}_{B}^0(k_0)$,
where the state for the environment $B$ is given by
\begin{equation}
 \label{eq:rho0B}
   \rB^0(k_0)=\frac{1}{N_{k_0}}\sum_{n_{k_0}}\ketbra{n_{k_0}}{n_{k_0}}\,,
\end{equation} 
i.e., only band $k_0$ is occupied. 

Under the above conditions and for $\Dd=0$ the subsystem $B$ acts as a thermalizing environment on the TLS: 
Subject to undisturbed Schr\"odinger evolution of the total system (i.e., without measurements), the TLS would relax 
into a state with inverse temperature $T^{-1}=\beta(k_0)$ given by~(\ref{eq:beta}) independently of its initial state
 $\op{\rho}(0)$~\cite{Gemmer2009, Breuer2006, Gemmer2006}.

In the following, we want to study the effect of external observation (by means of periodic measurements) for this 
quantum-thermodynamical resonant case as well as for the case of a detuned environment (i.e., $\dB\neq\dS$).

\section{Effective environment}
Without measurements, the environmental spins have to be in resonance with the TLS, i.e., the detuning has to vanish ($\Dd=0$)
in order to allow for a thermalization of the TLS. However, this no longer needs to be the case if we include 
periodic measurements.
We thus analyze the dynamics of the system with arbitrary detuning $\Dd$ disturbed by periodic measurements of the environmental 
magnetization, i.e., the system first evolves under pure Schr\"odinger dynamics for some time $\Dt$. Then a measurement of the 
environmental magnetization (or rather its energy) is executed, which means we perform a projection on one of the energy bands $k_1$.
The pertinent projection operator reads
\begin{equation}
 \Ptot{k_1}=\op{1}_S\otimes\P{k_1}\,
\end{equation} 
with
\begin{equation}
 \label{eq:projectionoperator}
 \P{k_1}=\sum_{n_{k_1}}\ketbra{n_{k_1}}{n_{k_1}}\,,
\end{equation} 
where we sum over all levels of the measured energy band $k_1$.
This leads to a state of the TLS given by
\begin{equation}
 \label{eq:rhocojump}
 \rS(1)=\frac{\Tr[B]{\Ptot{k_1}\rtotbef(1)}}{\Tr[B]{\P{k_1}\rBbef(1)}}\,,
\end{equation} 
which can be rewritten as
\begin{equation}
 \label{eq:rhocojump2}
 \rS(1)=\rSbef(1)+\frac{\Tr[B]{\Ptot{k_1} \op{C}_{SB}(1)}}{\Tr[B]{\P{k_1}\rBbef(1)}}\,,
\end{equation} 
where $\rSbef$ denotes the state of the TLS before the measurement and the second term is the so-called
co-jump~\cite{Granzow1998} caused by the correlations  $\op{C}_{SB}:=\rtotbef-\rSbef\otimes\rBbef$ 
between system and environment straight before the measurement. Since the measurement is incomplete (measurement of an energy band
and not of a single level), there may still be some correlations left after the measurement. However, the effect of these correlations 
is negligible in the case of weak interaction and small band width.
That is, the state of the total system after the measurement can be approximated by a product state. For simplicity, we introduce a 
further approximation: Since we are not interested in details like the occupation probabilities of the single levels within the
energy bands of the environment, we apply some kind of coarse graining by replacing the typically rather complicated state of 
the environment after the measurement by a simple one of the form~(\ref{eq:rho0B}). This can be done, provided the environment is
sufficiently large.
Thus, the new state of the total system after the measurement is approximated by 
\begin{equation}
 \rtot(1)\approx\rS(1)\otimes\rB^0(k_1)\,,
\end{equation} 
i.e., a form similar to the initial total state but with a new state $\rS(1)$ for the system and possibly a different energy band being
occupied for the environment (if a different energy band has been measured). 
This procedure can now be repeated periodically. In order to analyze what happens with the TLS due to these measurements, we first 
have to calculate the state of the TLS after a single measurement~(\ref{eq:rhocojump}). 
This can be done by means of perturbation theory~(see Appendix~\ref{App:derivation}), which leads to three possible states of the TLS
corresponding to the three possible measurement results:

In the case of measuring the same energy as in the previous measurement, we obtain the occupation probability of the ground state up
to second order in the interaction strength $\lambda$:
\begin{equation}
\label{eq:rho000expansion}
 \rc{00,0}(j)\approx\rc{00}(j-1)\left( 1-4\lambda^2\rc{11}(j-1)(\E^{-\beta\dB/2}-\E^{\beta\dB/2})\left( \sinA-\sinB\right)\right)  
\end{equation}
with
\begin{eqnarray}
 \sinA&:=&\frac{\sin^2\left(\frac{\Dd}{2}\Dt\right)}{\Dd^2}\,,\\
 \sinB&:=&\frac{\sin^2\left(\left(\dS+\frac{\Dd}{2}\right)\Dt\right)}{(2\dS+\Dd)^2}\,.
\end{eqnarray} 
In case of measuring one energy band higher than before, we get
\begin{equation}
\label{eq:rho00p}
 \rc{00,+}(j)\approx\frac{\rc{11}(j-1)\sinA}{\rc{11}(j-1)\sinA+\rc{00}(j-1)\sinB } \,,
\end{equation} 
whereas measuring one energy band lower leads to
\begin{equation}
\label{eq:rho00m}
 \rc{00,-}(j)\approx\frac{\rc{11}(j-1)\sinB}{\rc{11}(j-1)\sinB+\rc{00}(j-1)\sinA } \,.
\end{equation} 
For the off-diagonal elements we get accordingly
\begin{eqnarray}
 \label{eq:rho100expansion}
  \rc{10,0}(j)&\approx&\rc{10}(j-1)\Bigg[1+\lambda^2\Bigg(\left(\E^{-\beta\dB/2}\rc{00}(j-1)+\E^{\beta\dB/2}\rc{11}(j-1)\right)4\sinA\nonumber\\
&&+\left(\E^{\beta\dB/2}\rc{00}(j-1)+\E^{-\beta\dB/2}\rc{11}(j-1)\right)4\sinB\nonumber\\
&&-\left(\E^{-\beta\dB/2}+\E^{\beta\dB/2}\right)\nonumber\\
&&\times\left(\frac{1-\E^{\I\Dd\Dt}+\I \Dd\Dt}{\Dd^2}+\frac{1-\E^{\I(2\dS+\Dd)\Dt}+\I (2\dS+\Dd)\Dt}{(2\dS+\Dd)^2}\right)\Bigg)\Bigg]\,.\\
\label{eq:rho10p}
 \rc{10,+}(j)&\approx&\rho_{01}(j-1)\frac{1+\E^{2\I \dS \Dt}-2\E^{\I\dS \Dt}\cos((\dS+\Dd)\Dt)}{4(2\dS\Dd+\Dd^2)\left(\rc{11}(j-1)\sinA+\rc{00}(j-1)\sinB\right)}\,,\\
\label{eq:rho10m}
 \rc{10,-}(j)&\approx&\rho_{01}(j-1)\frac{1+\E^{2\I \dS \Dt}-2\E^{\I\dS \Dt}\cos((\dS+\Dd)\Dt)}{4(2\dS\Dd+\Dd^2)\left(\rc{11}(j-1)\sinB+\rc{00}(j-1)\sinA\right)}\,.
\end{eqnarray} 

As one can see, measuring a different band causes significant changes in the occupation probabilities of the TLS: For example, if $\rho_{00}$ was
close to $1$ at the previous measurement it will be almost $0$ after such a measurement, and vice versa. Thus, one cannot expect
that a single TLS will relax to a stable thermal attractor state as is the case without measurements in the quantum thermodynamical model.
Indeed, after sufficiently many measurements the TLS will always be in its ground or its excited state, i.e., passing through
some kind of quasi-classical trajectory.

Instead, one may ask what happens to the TLS in ensemble average.
Therefore, we first have to weight each possible outcome with the probability to obtain the corresponding measurement result.
These probabilities are given by~(\ref{eq:TrPkrhoB}):
\begin{equation}
  p_{k_{j}}(j)=\Tr[B]{\P{k_j}\rBbef(j)}\,,
\end{equation} 
which equals the occupation probabilities of the bands $k_{j}=\{k_{j-1},k_{j-1}+1,k_{j-1}-1\}$ before the measurement.

Thus, for a given energy band $k_{j-1}$ determined at the preceding measurement $j-1$, the probabilities for measuring one band higher, one band 
lower, or the same band, respectively, read
\begin{eqnarray}\label{eq:probabilitybandup}
 p_{+}(j)&\approx&4\lambda^2\E^{\beta\dB/2}\left(\rho_{11}(j-1)\sinA+\rho_{00}(j-1)\sinB\right)\,,\\
 \label{eq:probabilitybanddown} 
 p_{-}(j)&\approx&4\lambda^2\E^{-\beta\dB/2}\left(\rho_{00}(j-1)\sinA+\rho_{11}(j-1)\sinB\right)\,,\\
 \label{eq:probabilitybandequal}
 p_{0}(j)&\approx&1- p_{+}(j)- p_{-}(j)\,.
\end{eqnarray} 
This yields the ensemble average (indicated by an overbar) for the diagonal elements,
\begin{eqnarray}
\label{eq:rho00short}
 \avrc{00}(j)&=&p_0(j) \rc{00,0}(j)+p_+(j) \rc{00,+}(j)+ p_-(j) \rc{00,-}(j)\nonumber\\
 &\approx&\rc{00}(j-1)+4\lambda^2\Big[ \left( \E^{\beta\dB/2} \rc{11}(j-1)-\E^{-\beta\dB/2}\rc{00}(j-1)\right) \sinA\nonumber\\
&&+\left( \E^{-\beta\dB/2} \rc{11}(j-1)-\E^{\beta\dB/2}\rc{00}(j-1)\right)\sinB\Big]\, .
\end{eqnarray}
We can now iterate this
result to get the ensemble-averaged density matrix as a function of the number of measurements, i.e., the ensemble average after
measurement $j$ is used as initial state for measurement $j+1$ and so forth. The calculation for the off-diagonal elements can be found
in Appendix~\ref{app:offdiagonalelements}. It turns out that the off-diagonal elements vanish for almost any choice of the parameters
after sufficiently many measurements (an interesting exception is discussed below), i.e., the attractor state of the TLS will be a thermal state.
We note in passing that the ensemble average coincides with the infinite-time average (ergodicity).

\section{Effective relaxation controlled by $\Dt$ and $\Dd$}
To derive the analytical expression for $\avrc{00}(j)$, we first rewrite~(\ref{eq:rho00short}) as
\begin{equation}
\label{eq:averagederivation1}
\avrc{00}(j+1)-\avrc{00}(j)= -R \avrc{00}(j)+d
\end{equation} 
with
\begin{eqnarray}
 R&:=&8\lambda^2\cosh(\beta\dB/2)\left(\sinA + \sinB\right)\,,\\
 d&:=&4\lambda^2\left(\E^{\beta\dB/2}\sinA+ \E^{-\beta\dB/2}\sinB\right)\,.
\end{eqnarray} 
The left-hand side can be approximated by a derivative with respect to $j$, as long as $\avrc{00}$ changes only slowly 
with $j$, which is guaranteed by the weak coupling ($\lambda\ll1$).
Thus, we can approximate~(\ref{eq:averagederivation1}) by a differential equation of the form
\begin{equation}
 \dnd{\avrc{00}(j)}{j}=-R \avrc{00}(j)+d\,,
\end{equation} 
with the solution
\begin{equation}
\label{eq:exponentialrelaxrho00}
 \avrc{00}(j)=\left(\rc{00}(0)-\frac{d}{R}\right)\E^{-R j}+\frac{d}{R}\,,
\end{equation} 
This means that the ensemble of TLSs exponentially approaches an attractor state with increasing number $j$ of measurements.

This relaxation has some interesting properties: First, it is noteworthy that due to the measurements a relaxation becomes possible
not only in the resonant case (as without measurements) but also for $\dB\neq\dS$. However, the attractor state reached 
as well as the relaxation constant depend on the time $\Dt$ between the measurements and on the detuning $\Dd$.

Considering the relaxation constant
\begin{equation}
\label{eq:relaxationconstant}
 R=8\lambda^2\cosh(\beta\dB/2)\left(\frac{\sin^2\left(\frac{\Dd}{2}\Dt\right)}{\Dd^2} + \frac{\sin^2\left(\left(\dS+\frac{\Dd}{2}\right)\Dt\right)}{(2\dS+\Dd)^2}\right)\,,
\end{equation} 
one finds
\begin{equation}
 \lim_{\Dt\rightarrow0}R=0\,,
\end{equation} 
which means that the relaxation is slowed down for very rapidly repeated measurements. Such a suppression of decay due to fast, 
periodic measurements is well known as the so-called quantum Zeno effect~\cite{Misra1977,Itano1990}.
Furthermore, $R$ also vanishes for $\Dt=n\frac{\pi}{\dS}$ and $\Dd=\frac{2m\pi}{\Dt}$, $n=1, 2,\ldots$, $m=1, 2,\ldots$. 
For this special choice of the parameters (characterizing the exceptions we mentioned before), the absolute value of the off-diagonal elements also stays 
constant, as discussed in Appendix \ref{app:offdiagonalelements}.
Thus, it is possible to freeze the initial state of the TLS and suppress decoherence by periodic measurements carried out even 
with finite frequency.

For any other choice of the parameters, however, a thermal attractor state is obtained, which also shows remarkable features.
In general, it is given by (cf.~(\ref{eq:exponentialrelaxrho00}))
\begin{eqnarray}
 \label{eq:rho00Attr}
 \attr{00}&=&\frac{d}{R}\nonumber\\
 &=&\frac{\E^{\beta\dB/2} \sinA+\E^{-\beta\dB/2}\sinB }{2\cosh(\beta\dB/2)\left(\sinA + \sinB\right) }\,.
\end{eqnarray}
Let us first study the resonant case $\Dd=0$. Because of
\begin{eqnarray}
\lim_{\Dd\rightarrow0} \sinA&=&\lim_{\Dd\rightarrow0}\frac{\sin^2\left(\frac{\Dd}{2}\Dt\right)}{\Dd^2}=\frac{\Dt^2}{4}\,,\\
\lim_{\Dd\rightarrow0} \sinB&=&\lim_{\Dd\rightarrow0}\frac{\sin^2\left(\left(\dS+\frac{\Dd}{2}\right)\Dt\right)}{(2\dS+\Dd)^2}=\frac{\sin^2\left(\dS\Dt\right)}{4\dS^2}\,,
\end{eqnarray}
we get
\begin{equation}
 \label{eq:rho00AttrRes}
 \attr{00}(\Dt)=\frac{\E^{-\beta\dS/2}\sin^2(\dS \Dt)+ \E^{\beta\dS/2} \dS^2 \Dt^2}{2\cosh(\beta\dS/2)\left(\sin^2(\dS \Dt) + \dS^2 \Dt^2\right) }\,.
\end{equation}
The lowest possible temperature for the TLS in this resonant case is obtained for $\Dt=\frac{n\pi}{\dS}$, $n=1,2,3,\ldots$ and is given by
$T_{\text{min,res}}=1/\beta$. This is just the temperature the TLS would also get according to
quantum thermodynamics without measurements (cf.~\cite{Jahnke2010}). However, any other choice of $\Dt$ will lead to a higher temperature. In particular, for
rapidly repeated measurements we obtain 
\begin{equation}
  \label{eq:rho00AttrResLimit1}
  \lim_{\Dt\rightarrow0}\attr{00}=\frac{1}{2}\,,
\end{equation}
i.e., the TLS is heated up to very high temperatures ($T_\text{eff}\rightarrow\infty$).

Finally we want to focus on the effect of a detuning $\Dd\neq0$ between system and environment: Now, the 
attractor state~(\ref{eq:rho00Attr}) reads
\begin{eqnarray}
 \label{eq:rho00AttrOffres}
 \attr{00}&=&\frac{\E^{\beta\dB/2} \frac{\sin^2\left(\frac{\Dd}{2}\Dt\right)}{\Dd^2}+\E^{-\beta\dB/2}\frac{\sin^2\left(\left(\dS+\frac{\Dd}{2}\right)\Dt\right)}{(2\dS+\Dd)^2} }{2\cosh(\beta\dB/2)\left(\frac{\sin^2\left(\frac{\Dd}{2}\Dt\right)}{\Dd^2} + \frac{\sin^2\left(\left(\dS+\frac{\Dd}{2}\right)\Dt\right)}{(2\dS+\Dd)^2}\right) }\,,
\end{eqnarray}
and has two new interesting properties compared to the resonant case discussed before.
The lowest possible temperature here is reached for $\Dt=n\pi/(\dS+\Dd/2)$, $n=1, 2, 3,\ldots$, and is given by
\begin{equation}
 T_\text{min}=\frac{\dS}{\dB}\frac{1}{\beta}\,.
\end{equation} 
This temperature depends on the ratio of the energy splittings of the TLS and environmental spins,
that is, the off-resonance may be used for cooling the TLS down to very low temperatures by choosing proper values
for $\Dt$ and $\Dd$.

The other remarkable feature is the limit of the lowest possible occupation probability of the ground state, i.e., the highest temperature.
As in the resonant case, rapidly repeated measurements lead to $T_\text{eff}\rightarrow\infty$. However, in the off-resonant case, it is even
possible to further increase the energy of the TLS, i.e., to create inversion. The maximum inversion is reached for $\Dt=2 n\pi/|\Omega|$, $n=1, 2, 3,\ldots$, and is given by
\begin{equation}
  \label{eq:rho00AttrOffresLimit2}
  \text{min}\left(\attr{00}(\Dt)\right)=\frac{\E^{-\beta\dB/2}}{\E^{\beta\dB/2}+\E^{-\beta\dB/2}}\,.
\end{equation}
which corresponds to the negative temperature~\cite{Landsberg1959}
\begin{equation}
 T_\text{max}=-\frac{\dS}{\dB}\frac{1}{\beta}\,.
\end{equation} 
Note, that this inversion originates from the influence of the periodic measurements and not from a negative temperature state
of the spin environment (the environmental $\beta$ is positive).

\begin{figure}
\begin{center}
\includegraphics[scale=0.8]{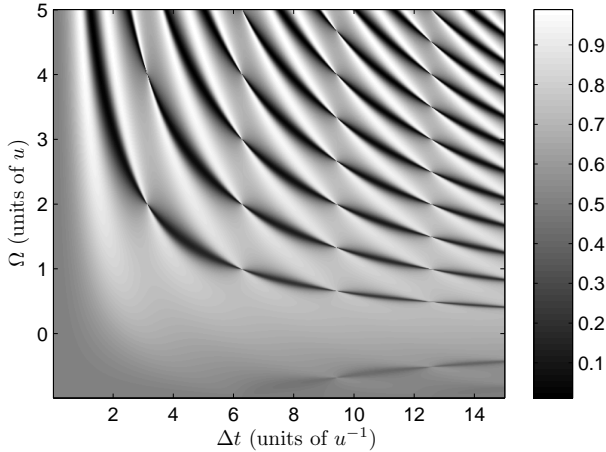}
\end{center}
\caption{Attractor for the occupation probability of the ground state $\attr{00}$ as a function of the parameters $\Dt$ and $\Dd$. States with low 
temperature are white and those with high (negative) temperature are black. Parameters used: $\delta=1\eu$, $\beta=0.75\eu^{-1}$. \label{fig:attractorOffres}}
\end{figure}

By varying the time $\Dt$ between the measurements, thermal states of arbitrary temperature within these limits can be prepared.
Possible values of inverse temperatures thus cover the range between $-(\dB/\dS) \beta$ and $+(\dB/\dS) \beta$, symmetrically around
zero. (In the resonant case these values are between zero and $\beta$ only.)
Fig.~\ref{fig:attractorOffres} shows the attractor for the occupation probability of the ground state as a function of time $\Dt$ and
detuning $\Dd$.

\section{Numerical verification for small environments}
The discussed behavior of the TLS under periodic measurements of the environment can be obtained not only for large environments but even
for quite small ones consisting of a few spins only. For such small environments the attractor state for the TLS
depends not only on the initial state of the environment (working point) but also on the initial state of the system, since the 
environmental degeneracy structure can no longer be approximated by an exponential behavior. This means that the system no 
longer shows thermodynamic behavior, which requires independence of its initial state. Nevertheless, given the initial state, we
can predict the attractor state using our analytical results.

\begin{figure}
\begin{center}
\includegraphics[scale=0.7]{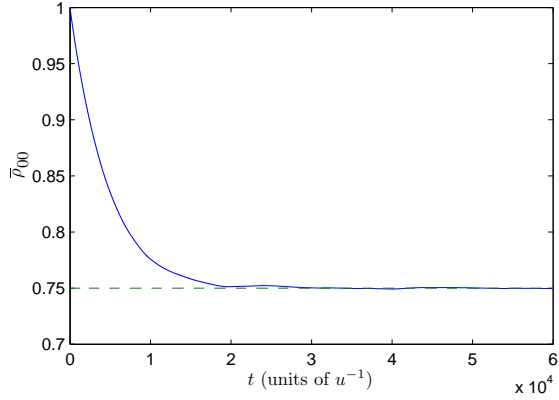}
\end{center}
\caption{(Color online) Relaxation of the TLS given by the ensemble-averaged $\avrc{00}$ under periodic measurements of its environment
(seven spins with initially two spins up).
We obtain a relaxation to the thermal attractor state with $\attr{00}=3/4$ (dashed line).
Chosen parameters: $\dS=1\eu$, $\Dd=0$, $\Dt=\pi/\delta$. \label{fig:rho00measdtpiquasiattr}}
\end{figure}

To demonstrate this, we consider a TLS with splitting $\dS$ coupled via a random $(\sigma_x\otimes\sigma_x)$-interaction 
to a seven-spin environment (splitting of the spins $\dB$), with initially two spins up. 
Let us first consider the resonant case $\dS=\dB=1\eu$ with $\Dt=\pi/\dS$ and the TLS initially being in its ground state. 
In this case only the initially occupied environmental ``energy band'' $E_2=2\dB$ and the band beneath ($E_1=\dB$) are involved 
in the dynamics (the ground state energy is set to $E_0=0$). This means that only the ratio of the degeneracies of these bands determines the final attractor state.
Thus, because $N_1={{7}\choose{1}}=7$ and $N_2={{7}\choose{2}}=21$, we get an effective inverse temperature 
$\beta_\text{eff}=\ln(N_2/N_1)\dB^{-1}=\ln(3)\dB^{-1}$ and therefore expect $\attr{00}=3/4$. Indeed, this final state is obtained in
our numerical exact simulation shown in Fig.~\ref{fig:rho00measdtpiquasiattr}.

\begin{figure}
\begin{center}
\includegraphics[scale=0.7]{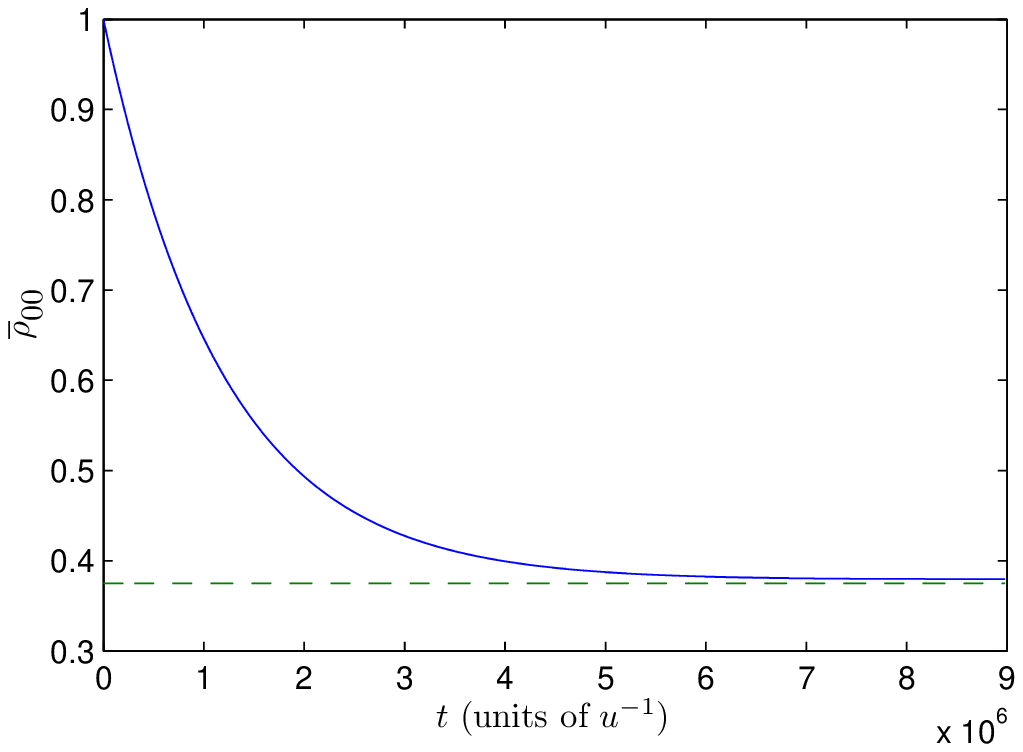}
\end{center}
\caption{(Color online) Relaxation of the TLS to a state of negative temperature due to the periodic measurements of its environment (seven spins
with initially two spins up). The final state is determined by $\attr{00}=3/8$ (dashed green line). Chosen parameters: $\dS=1\eu$, $\Dd=0.7\eu$,
$\Dt=2\pi/\Omega$. \label{fig:rho00measdt2pioverOmega}}
\end{figure}

In the off-resonant case, it is possible to obtain inversion as discussed in the last section. To illustrate this, we choose
$\Dd=0.7$ and $\Dt=2\pi/\Dd$. Again, initially we start with two spins up. Since now only the second and third bands are involved, we 
obtain $\beta_\text{eff}=\ln(35/21)\dB^{-1}$, which yields according to~(\ref{eq:rho00AttrOffresLimit2})
$\attr{00}=3/8$. Again, this value is verified by the simulation, as shown in Fig.~\ref{fig:rho00measdt2pioverOmega}.

\section{Conclusion}
In this paper, we have studied the influence of an effective environment on a coupled two-level
system. We have shown analytically that periodic measurements of the environmental magnetization (energy) may lead to a freezing of the initial state, i.e., suppress decoherence,
or may (in ensemble average) yield a thermal state for the TLS, depending on the choice of the detuning and the time between the measurements. The 
temperature of the attractor state reached is controlled by these two parameters, which allows for the preparation of thermal
states with almost arbitrary temperature. It is even possible to create an inversion for the TLS. The key mechanism, which underlies
this cooling or creation of inversion, is the periodic destruction of the correlations between the system and its detuned 
environment. The resulting periodic reset of the interaction energy allows for a successive change of energy of the system and 
environment, even if out of resonance. It is to be expected that the deviation $T_\text{eff}$ from $T$ would give way to non-thermal
attractor states in the case of larger subsystems $S$, i.e., beyond the two-level case.~\footnote{The potential generalization to 
larger systems poses a challenging question, which should be addressed in future work. We expect that, in general, periodic
measurements of a thermalizing environment will only lead to the undisturbed thermal state of the system if 
the repetition of measurements is sufficiently slow, such that the effect of the periodic reset of the interaction energy
is negligible.}

As demonstrated in our numerical simulations, this intriguing behavior can be observed even for quite small environments consisting
of a few spins only. In this case, the attractor exhibits dependence on the initial state of the system $S$.

In a recent paper~\cite{Alvarez2010} a nuclear magnetic resonance setup has been used for studying the influence of periodic 
quantum nondemolition measurements in a related model. We think that similar experiments may also be suitable to experimentally
test the results we have derived here.

\appendix

\section{Thermalizing environment\label{sec:thermalizingenvironment}}
In this appendix, we show how the degeneracy structure of a system consisting of many spins can be approximated by an exponentially
increasing degeneracy around a certain point in energy space, which is a  precondition for this system to act as a thermalizing
environment~\cite{Gemmer2009}.
Thus, we want to approximate
\begin{equation}
 {{n}\choose{k}}\approx\mathcal{N}\E^{\beta\dB k}
\end{equation} 
around some working point $k_0$ (we set the ground state energy to $0$).
To do so, we first replace the discrete binomial coefficient by a continuous function, which can be done using $n!=\Gamma(n+1)$, with
$\Gamma(x)$ being the Gamma function. 
Thus, the binomial coefficient can be written as
\begin{equation}
 {{n}\choose{k}}=\frac{\Gamma(n+1)}{\Gamma(n-k+1)\Gamma(k+1)}\,.
\end{equation} 
Now, we can apply a Taylor expansion around $k_0$, which reads
\begin{equation}
  {{n}\choose{k}}=\frac{\Gamma(n+1)}{\Gamma(n-k_0+1)\Gamma(k_0+1)}\left(1+(\psi(n-k_0+1)-\psi(k_0+1))(k-k_0)\right)+\mathcal{O}\left((k-k_0)^2\right)
\end{equation} 
where $\psi(x):=\frac{\D}{\D x}\ln(\Gamma(x))$ is the so-called Digamma function~\cite{Anderson1997}.
Comparison of this with the Taylor expansion of $\mathcal{N}\E^{\beta\dB k}$ yields in zeroth order
\begin{equation}
\label{eq:binomialtaylor0}
 \frac{n!}{k_0!(n-k_0)!}\stackrel{!}{=}\mathcal{N}\E^{\beta\dB k_0}
\end{equation} 
and in first order
\begin{equation}
\label{eq:binomialtaylor1}
 \frac{n!}{k_0!(n-k_0)!}(\psi(n-k_0+1)-\psi(k_0+1))\stackrel{!}{=}\mathcal{N}\beta\dB\E^{\beta\dB k_0}\,.
\end{equation} 
Hence, with~(\ref{eq:binomialtaylor1})/(\ref{eq:binomialtaylor0}) we get
\begin{equation}
 \beta\dB=\psi(n-k_0+1)-\psi(k_0+1)
\end{equation} 
For large $x$ the Digamma function can be approximated by $\psi(x)\approx\ln(x)$, which finally leads to equation~(\ref{eq:beta}), i.e.,
\begin{equation}
\label{eq:betaspinumgebung}
 \beta\approx\frac{1}{\dB}\ln\left(\frac{n}{k_0}-1\right).
\end{equation} 
This means, that such an environment consisting of many spins with total energy $\dB k_0$ can act as a thermalizing environment 
on a coupled TLS with $\dS=\dB$, enforcing a thermal state of temperature $1/\beta$ on the TLS according to~(\ref{eq:betaspinumgebung}).

\section{Calculation of the short-time dynamics\label{App:derivation}}
The von Neumann equation for the total system
in the interaction picture (we set $\hbar=1$) reads
\begin{equation}
 \label{eq:vonNeumanneq}
 \frac{\partial}{\partial t}\rtot(t) = \I \kom{\rtot(t)}{\lambda \op{V}(t)}
\end{equation} 
with
\begin{equation}
\label{eq:Vt}
 \op{V}(t)=\op{\sigma}^{+}\op{B}(t)+\op{\sigma}^{-}\op{B}^\dagger(t)
\end{equation} 
and
\begin{eqnarray}
 \op{B}(t)&=&\E^{\I \op{H}_B t}\op{B}\E^{-\I \op{H}_B t}\E^{\I \dS t}\nonumber\\
&=&\sum_{k}\sum_{n_k,m_{k+1}}C_{k+1,k}(m_{k+1},n_{k})\E^{-\I(\Dd+\omega(m_{k+1},n_{k}))t}\ketbra{n_k}{m_{k+1}}\nonumber\\
&&+C_{k,k+1}(n_k,m_{k+1})\E^{\I(2\delta+\Dd+\omega(m_{k+1},n_{k}))t}\ketbra{m_{k+1}}{n_k}\,,
\end{eqnarray} 
where $\omega(m_{k+1},n_{k})$ is the energy difference of level $m$ in band $k+1$ and level $n$ in band $k$ minus $\dS$, i.e., always
much smaller than $\dS$ for small band width.

Since our goal is to study the influence of the periodic measurements, we denote the state of the system by the discrete number of the
measurement instead of the continuous time, i.e., $\rS(j)\equiv\rS(t=j\Dt)$ and so forth.
According to~(\ref{eq:vonNeumanneq}), the density operator of the total system up to second order in interaction strength
can then be written as
\begin{eqnarray}
 \label{eq:expansionrhotot}
 \rtotbef(j)&\approx&\rtot(j-1)+\I\lambda\int_0^\Dt\kom{\rtot(j-1)}{\op{V}(t^\prime)}\D t^\prime\nonumber\\
&&-\lambda^2\int_0^\Dt\int_0^{t^\prime}\kom{\kom{\rtot(j-1)}{\op{V}(t^{\prime\prime})}}{\op{V}(t^\prime)}\D t^{\prime\prime} \D t^\prime\,.
\end{eqnarray} 
One can easily show that the first-order term does not 
contribute to the state of the TLS since $\Tr[B]{\rB(j-1)\op{B}(t^\prime)}=\Tr[B]{\rB(j-1)\op{B}^\dagger(t^\prime)}=0$.

The commutator appearing in the second-order term reads
\begin{eqnarray}
\label{eq:2ndordercommutator}
 &&\!\!\!\!\!\!\!\kom{\kom{\rtot(j-1)}{\op{V}(t^{\prime\prime})}}{\op{V}(t^\prime)}=\nonumber\\
&&\rtot(j-1)\op{\sigma}^{+}\op{B}(t^{\prime\prime})\op{\sigma}^{-}\op{B}^\dagger(t^{\prime}) + \rtot(j-1)\op{\sigma}^{-}\op{B}^\dagger(t^{\prime\prime})\op{\sigma}^{+}\op{B}(t^{\prime})\nonumber\\
&&-\op{\sigma}^{+}\op{B}(t^{\prime\prime})\rtot(j-1)\op{\sigma}^{+}\op{B}(t^{\prime}) - \op{\sigma}^{+}\op{B}(t^{\prime\prime})\rtot(j-1)\op{\sigma}^{-}\op{B}^\dagger(t^{\prime})\nonumber\\
&&-\op{\sigma}^{-}\op{B}^\dagger(t^{\prime\prime})\rtot(j-1)\op{\sigma}^{+}\op{B}(t^{\prime}) - \op{\sigma}^{-}\op{B}^\dagger(t^{\prime\prime})\rtot(j-1)\op{\sigma}^{-}\op{B}^\dagger(t^{\prime})\nonumber\\
&&-\op{\sigma}^{+}\op{B}(t^{\prime})\rtot(j-1)\op{\sigma}^{+}\op{B}(t^{\prime\prime}) - \op{\sigma}^{-}\op{B}^\dagger(t^{\prime})\rtot(j-1)\op{\sigma}^{+}\op{B}(t^{\prime\prime})\nonumber\\
&&-\op{\sigma}^{+}\op{B}(t^{\prime})\rtot(j-1)\op{\sigma}^{-}\op{B}^\dagger(t^{\prime\prime}) - \op{\sigma}^{-}\op{B}^\dagger(t^{\prime})\rtot(j-1)\op{\sigma}^{-}\op{B}^\dagger(t^{\prime\prime})\nonumber\\
&&+\op{\sigma}^{-}\op{B}^\dagger(t^{\prime})\op{\sigma}^{+}\op{B}(t^{\prime\prime})\rtot(j-1) + \op{\sigma}^{+}\op{B}(t^{\prime})\op{\sigma}^{-}\op{B}^\dagger(t^{\prime\prime})\rtot(j-1)\nonumber\\
&&=: \secordcomm\,.
\end{eqnarray}

This leads to the $00$ component of the numerator $\Tr[B]{\Ptot{k_j}\rtotbef(j)}$ in~(\ref{eq:rhocojump})
\begin{eqnarray}
\label{eq:TrBPkrho}
 \Tr[B]{\Ptot{k_j}\rtotbef(j)}_{00}&\approx&\rc{00}(j-1)\delta_{k_{j-1} k_{j}}\nonumber\\
&&-\lambda^2\int_0^\Dt\int_0^{t^\prime}\rc{00}(j-1)\xi_0(t^\prime,t^{\prime\prime})\nonumber\\
&&\quad\quad\quad\quad\quad-\rc{11}(j-1)\xi_1(t^\prime,t^{\prime\prime}) \D t^{\prime\prime}\D t^\prime
\end{eqnarray}
with
\begin{eqnarray}
 \xi_0(t^\prime,t^{\prime\prime})&:=&\Tr[B]{\P{k_j}\rB(j-1)\op{B}^\dagger(t^{\prime\prime})\op{B}(t^\prime) + \P{k_j}\op{B}^\dagger(t^{\prime})\op{B}(t^{\prime\prime})\rB(j-1)}\nonumber\\
&=&\frac{2\delta_{k_{j-1}k_{j}}}{N_{k_j}}\sum_{m_{k_j},n_{k_j-1}}|C_{k_j,k_j-1}(m_{k_j},n_{k_j-1})|^2\nonumber\\
&&\quad\quad\quad\quad\quad\quad\quad\quad\times\cos((\Dd+\omega(m_{k_j},n_{k_j-1}))(t^\prime-t^{\prime\prime}))\nonumber\\
&&+\frac{2\delta_{k_{j-1}k_j}}{N_{k_j}}\sum_{m_{k_j},n_{k_j+1}}|C_{k_j,k_j+1}(m_{k_j},n_{k_j+1})|^2\nonumber\\
&&\quad\quad\quad\quad\quad\quad\quad\quad\times\cos((2\delta+\Dd+\omega(n_{k_j+1},m_{k_j}))(t^\prime-t^{\prime\prime}))\,,\nonumber\\
\xi_1(t^\prime,t^{\prime\prime})&:=&\Tr[B]{\P{k_j}\op{B}^\dagger(t^{\prime\prime})\rB(j-1)\op{B}(t^\prime) + \P{k_j}\op{B}^\dagger(t^{\prime})\rB(j-1)\op{B}(t^{\prime\prime})}\nonumber\\
&=&\frac{2\delta_{k_{j-1}+1 k_{j}}}{N_{k_j-1}}\sum_{m_{k_j},n_{k_j-1}}|C_{k_j,k_j-1}(m_{k_j},n_{k_j-1})|^2\nonumber\\
&&\quad\quad\quad\quad\quad\quad\quad\quad\times\cos((\Dd+\omega(m_{k_j},n_{k_j-1}))(t^\prime-t^{\prime\prime}))\nonumber\\
&&+\frac{2\delta_{k_{j-1}-1k_j}}{N_{k_j+1}}\sum_{m_{k_j},n_{k_j+1}}|C_{k_j,k_j+1}(m_{k_j},n_{k_j+1})|^2\nonumber\\
&&\quad\quad\quad\quad\quad\quad\quad\quad\times\cos((2\delta+\Dd+\omega(n_{k_j+1},m_{k_j}))(t^\prime-t^{\prime\prime}))\,,
\end{eqnarray}
where $\delta_{ab}$ denotes the Kronecker delta. 
$\xi_0$ and $\xi_1$ both consist of two double sums over many oscillating terms with normally distributed amplitudes and approximately
uniformly distributed frequencies. For sufficiently small width of the bands (or rather short enough times) these sums can be
approximated by
\begin{equation}
\label{eq:shorttimeapprox}
 \sum_{m_{a},n_{b}}|C_{a,b}(m_{a},n_{b})|^2\cos((\alpha+\omega(m_{a},n_{b}))(t^\prime-t^{\prime\prime}))\approx N_a N_b \overline{|C_{a,b}(m_a,n_b)|^2}\cos(\alpha(t^\prime-t^{\prime\prime}))
\end{equation} 
with $\alpha$ being $\Dd$ or $2\dS+\Dd$, respectively.
Using $\overline{|C_{i,j}(a_i,b_j)|^2}=\left(N_i N_j\right)^{-1/2}=\E^{-\beta\dB(i+j)/2}$, we obtain
\begin{eqnarray}
 \xi_0(t^\prime,t^{\prime\prime})&\approx&2\delta_{k_{j-1}k_{j}}\big(\E^{-\beta\dB/2}\cos(\Dd(t^\prime-t^{\prime\prime}))\nonumber\\
&&+\E^{\beta\dB/2}\cos((2\dS+\Dd)(t^\prime-t^{\prime\prime}))\big)\,,\nonumber\\
 \xi_1(t^\prime,t^{\prime\prime})&\approx&2\delta_{k_{j-1}+1 k_{j}}\E^{\beta\dB/2} \cos(\Dd(t^\prime-t^{\prime\prime}))\nonumber\\
&&+ 2\delta_{k_{j-1}-1k_j}\E^{-\beta\dB/2}\cos((2\delta+\Dd)(t^\prime-t^{\prime\prime}))\,.\nonumber\\
\end{eqnarray} 
With this, the integrals in (\ref{eq:TrBPkrho}) can be easily calculated, leading to
\begin{eqnarray}
\label{eq:TrBPkrhofinal}
 \Tr[B]{\Ptot{k_j}\rtotbef}_{00}&\approx&\rc{00}(j-1)\delta_{k_{j-1} k_{j}}\nonumber\\
&&+4\lambda^2\Bigg[-\rc{00}(j-1)\delta_{k_{j-1}k_j}\left(\E^{-\beta\dB/2}\sinA+\E^{\beta\dB/2}\sinB\right)\nonumber\\
&&+\rc{11}(j-1)\nonumber\\
&&\quad\times\left(\E^{\beta\dB/2}\sinA\delta_{k_{j-1}+1 k_{j}}+\E^{-\beta\dB/2}\sinB\delta_{k_{j-1}-1 k_{j}}\right)\Bigg]\,,\nonumber\\
\end{eqnarray}
where we have introduced the abbreviations
\begin{eqnarray}
 \sinA&:=&\frac{\sin^2\left(\frac{\Dd}{2}\Dt\right)}{\Dd^2}\,,\\
 \sinB&:=&\frac{\sin^2\left(\left(\dS+\frac{\Dd}{2}\right)\Dt\right)}{(2\dS+\Dd)^2}\,.
\end{eqnarray} 

The off-diagonal element $\Tr[B]{\op{P}_k\op{\rho}_\text{tot}}_{10}$ can be calculated in the same way, leading to
\begin{eqnarray}
\label{eq:TrBPkrho10}
 \Tr[B]{\Ptot{k_j}\rtotbef(j)}_{10}&\approx&\rc{10}(j-1)\delta_{k_{j-1} k_{j}}\nonumber\\
&&-\lambda^2\int_0^\Dt\int_0^{t^\prime}\rc{10}(j-1)\text{Tr}_{B}\Big\{\P{k_j}\rB(j-1)\op{B}^\dagger(t^{\prime\prime})\op{B}(t^\prime)\nonumber\\
&&\quad\quad\quad\quad\quad\quad\quad\quad\quad\quad\quad\quad + \P{k_j}\op{B}(t^{\prime})\op{B}^\dagger(t^{\prime\prime})\rB(j-1)\Big\}\nonumber\\
&&\quad\quad\quad\quad\quad-\rc{01}(j-1))\text{Tr}_{B}\Big\{\P{k_j}\op{B}(t^{\prime\prime})\rB(j-1)\op{B}(t^\prime)\nonumber\\
&&\quad\quad\quad\quad\quad\quad\quad\quad\quad\quad\quad\quad\,\, + \P{k_j}\op{B}(t^{\prime})\rB(j-1)\op{B}(t^{\prime\prime})\Big\} \D t^{\prime\prime}\D t^\prime\nonumber\\
&\approx&\rc{10}(j-1)\delta_{k_{j-1}k_j}-\lambda^2\bigg[\delta_{k_{j-1}k_j}\left(\E^{\beta\dB/2}+\E^{-\beta\dB/2}\right)\nonumber\\
&&\times\!\left(\frac{1-\E^{\I\Dd \Dt}+\I\Dd \Dt}{\Dd^2}+\frac{1-\E^{\I(2\dS+\Dd) \Dt}+\I(2\dS+\Dd) \Dt}{(2\dS+\Dd)^2}\right)\nonumber\\
&&\times\rc{10}(j-1)\nonumber\\
&&+\frac{1+\E^{2\I\dS \Dt}(1-2\cos((\dS+\Dd)\Dt))}{2\dS\Dd+\Dd^2}\nonumber\\
&&\times\!\left(\delta_{k_{j-1}-1k_j}\E^{-\beta\dB/2}+\delta_{k_{j-1}+1 k_{j}}\E^{\beta\dB/2}\right)\nonumber\\
&&\times\rc{01}(j-1)\bigg]\,.
\end{eqnarray} 

To get the density matrix of the TLS, we finally have to calculate the denominator in~(\ref{eq:rhocojump}), which yields
\begin{eqnarray}
\label{eq:TrPkrhoB}
 \Tr[B]{\P{k_j}\rBbef(j)}&\approx&\delta_{k_{j-1}k_j}-\lambda^2\int_0^\Dt\int_0^{t^\prime}\Tr[B]{\P{k_j}\Tr[S]{\secordcomm}}\D t^{\prime\prime} \D t^\prime\nonumber\\
&\approx&\delta_{k_{j-1}k_j}-4\lambda^2\Bigg\lbrace\bigg[\rho_{11}(j-1)\left(\E^{\beta\dB/2}\sinA+\E^{-\beta\dB/2}\sinB\right)\nonumber\\
&&+\rho_{00}(j-1)\left((\E^{-\beta\dB/2}\sinA+\E^{\beta\dB/2}\sinB\right)\bigg]\delta_{k_{j-1}k_j}\nonumber\\
&&-\E^{\beta\dB/2}\left(\rho_{11}(j-1)\sinA+\rho_{00}(j-1)\sinB\right)\delta_{k_{j-1}+1 k_j}\nonumber\\
&&-\E^{-\beta\dB/2}\left(\rho_{00}(j-1)\sinA+\rho_{11}(j-1)\sinB\right)\delta_{k_{j-1}-1 k_j}\Bigg\rbrace\,.\nonumber\\
\end{eqnarray}
Equations~(\ref{eq:TrBPkrhofinal}), (\ref{eq:TrBPkrho10}), and (\ref{eq:TrPkrhoB}) thus lead to three different values for
 $\rc{00}(j)$ and $\rc{10}(j)$ after the measurement (equations~(\ref{eq:rho000expansion})--(\ref{eq:rho10m})),
depending on whether a lower, the same, or a higher energy has been measured compared to the previous measurement.

\section{Dynamics of the off-diagonal elements \label{app:offdiagonalelements}}
In this appendix we derive the dynamics of the off-diagonal elements of the TLS state due to the periodic measurements.
Using (\ref{eq:rho100expansion})--(\ref{eq:rho10m}) as well as (\ref{eq:probabilitybandup})--(\ref{eq:probabilitybandequal}),
we find for the ensemble average
\begin{eqnarray}
\label{eq:rho10short}
 \avrc{10}(j)&\approx&\avrc{10}(j-1)+2\lambda^2\cosh(\beta\dB/2)\nonumber\\
&&\times\bigg[-\avrc{10}(j-1)\nonumber\\
&&\quad\times\bigg(\frac{1-\E^{\I\Dd\Dt}+\I\Dd\Dt}{\Dd^2}+\frac{1-\E^{\I(2\dS+\Dd)\Dt}+\I(2\dS+\Dd)\Dt}{(2\dS+\Dd)^2}\bigg)\nonumber\\
&&\quad+\avrc{01}(j-1)\frac{1+\E^{2\I\dS\Dt}-2\E^{\I\dS\Dt}\cos((\dS+\Dd)\Dt)}{2\dS\Dd+\Dd^2} \bigg]\,,
\end{eqnarray}
i.e., the dynamics of the off-diagonal elements can be treated independently of the diagonal elements.
We first rewrite (\ref{eq:rho10short}) as
\begin{equation}
\label{eq:averagerho01derivation1}
 \avrc{10}(j)-\avrc{10}(j-1)=\avrc{10}(j-1)(c_1+\I c_2) + \avrc{01}(j-1)(c_3+\I c_4)
\end{equation} 
with
\begin{eqnarray}
\label{eq:defc1}
 c_1&:=&-2\lambda^2\cosh(\beta\dB/2)\!\left(\frac{1-\cos(\Dd\Dt)}{\Dd^2}+\frac{1-\cos((2\dS+\Dd)\Dt)}{(2\dS+\Dd)^2}\right)\!\!,\\
 c_2&:=&-2\lambda^2\cosh(\beta\dB/2)\nonumber\\
 \label{eq:defc2} 
&&\times\left(\frac{\Dd\Dt-\sin(\Dd\Dt)}{\Dd^2}+\frac{(2\dS+\Dd)\Dt-\sin((2\dS+\Dd)\Dt)}{(2\dS+\Dd)^2}\right),\\
 \label{eq:defc3}
 c_3&:=&-4\lambda^2\cosh(\beta\dB/2)\frac{\cos(\dS\Dt)\cos((\dS+\Dd)\Dt)-\cos^2(\dS\Dt)}{2\dS\Dd+\Dd^2}\,,\\
 \label{eq:defc4} 
 c_4&:=&-2\lambda^2\cosh(\beta\dB/2)\frac{2\sin(\dS\Dt)\cos((\dS+\Dd)\Dt)-\sin(2\dS\Dt)}{2\dS\Dd+\Dd^2}\,.
\end{eqnarray} 
Due to the weak coupling, we can approximate the left-hand side of~(\ref{eq:averagerho01derivation1}) by a derivative, which leads us to two coupled 
differential equations for $\ReOff(j):=\Re{\avrc{10}(j)}$ and  $\ImOff(j):=\Im{\avrc{10}(j)}$:
\begin{eqnarray}
\dnd{\ReOff(j)}{j}&=&(c_1+c_3)\ReOff(j)+(c_4-c_2)\ImOff(j)\,,\\
\dnd{\ImOff(j)}{j}&=&(c_1-c_3)\ImOff(j)+(c_2+c_4)\ReOff(j)\,.
\end{eqnarray}
The solution of this system of differential equations reads
\begin{eqnarray}
 \ReOff(j)&=&\frac{1}{2\gam}\E^{(c_1-\gam)j}\Big[\Big(c_3\left(\E^{2\gam j}-1\right)+\gam\left(\E^{2\gam j}+1\right)\Big)\ReOff(0)\nonumber\\
&&+(c_4-c_2)\left(\E^{2\gam j}-1\right)\ImOff(0)\Big]\,,\\
\ImOff(j)&=&\frac{1}{2\gam}\E^{(c_1-\gam)j}\Big[\left(c_2+c_4\right)\left(\E^{2\gam j}-1\right)\ReOff(0)\nonumber\\
&&+\left(-c_3\left(\E^{2\gam j}-1\right)+\gam\left(\E^{2\gam j}+1\right)\right)\ImOff(0)\Big]\,,
\end{eqnarray}  
where we introduced the abbreviation $\gam:=\sqrt{-c_2^2+c_3^2+c_4^2}$. 
This finally yields the absolute value of $\avrc{10}(j)$ 
\begin{eqnarray}
\label{eq:AbsOffdiagonal}
 |\avrc{10}(j)|&=&\left(\ReOff(j)^2+\ImOff(j)^2\right)^{1/2}\nonumber\\
&=&\frac{\E^{c_1 j}}{\gam}\Big[\big(\left(c_3\sinh(\gam j)+\gam\cosh(\gam j)\right)\ReOff(0)+(c_4-c_2)\sinh(\gam j)\ImOff(0)\big)^2\nonumber\\
&&+\big((c_2+c_4)\sinh(\gam j)\ReOff(0)+\left(-c_3\sinh(\gam j)+\gam\cosh(\gam j)\right)\ImOff(0)\big)^2\Big]^{1/2}\!\!\!.\nonumber\\
\end{eqnarray} 

Considering the special case $\Dt=n\frac{\pi}{\dS}$ and $\Dd=\frac{2m\pi}{\Dt}$,
 $n=1,2,\ldots$, $m=1,2,\ldots$, we find that $c_2=-\lambda^2\cosh(\beta\dB/2)\frac{n^2 (2m+n)\pi}{m(m+n)\delta^2}$ and $c_{1,3,4}=0$.
This directly leads us to
\begin{equation}
 \avrc{10}(j)=\avrc{10}(0)\E^{\I c_2 j}\,,
\end{equation} 
i.e., a slow oscillation of the off-diagonals.

Thus, the absolute value $|\avrc{10}(j)|$ stays constant for this special case, for which also the relaxation constant 
$R$~(\ref{eq:relaxationconstant}) vanishes. For any other choice of the parameters, we obtain from~(\ref{eq:AbsOffdiagonal})
$\lim_{j\rightarrow\infty}|\avrc{10}(j)|=0$, i.e., the TLS relaxes into a thermal state.


\begin{thebibliography}{15}
\expandafter\ifx\csname natexlab\endcsname\relax\def\natexlab#1{#1}\fi
\expandafter\ifx\csname bibnamefont\endcsname\relax
  \def\bibnamefont#1{#1}\fi
\expandafter\ifx\csname bibfnamefont\endcsname\relax
  \def\bibfnamefont#1{#1}\fi
\expandafter\ifx\csname citenamefont\endcsname\relax
  \def\citenamefont#1{#1}\fi
\expandafter\ifx\csname url\endcsname\relax
  \def\url#1{\texttt{#1}}\fi
\expandafter\ifx\csname urlprefix\endcsname\relax\def\urlprefix{URL }\fi
\providecommand{\bibinfo}[2]{#2}
\providecommand{\eprint}[2][]{\url{#2}}

\bibitem[{\citenamefont{Gemmer et~al.}(2009)\citenamefont{Gemmer, Michel, and
  Mahler}}]{Gemmer2009}
\bibinfo{author}{\bibfnamefont{J.}~\bibnamefont{Gemmer}},
  \bibinfo{author}{\bibfnamefont{M.}~\bibnamefont{Michel}}, \bibnamefont{and}
  \bibinfo{author}{\bibfnamefont{G.}~\bibnamefont{Mahler}},
  \emph{\bibinfo{title}{Quantum {T}hermodynamics}}
  (\bibinfo{publisher}{Springer, Berlin, Heidelberg}, \bibinfo{year}{2009}),
  \bibinfo{edition}{2nd} ed.

\bibitem[{\citenamefont{Misra and Sudarshan}(1977)}]{Misra1977}
\bibinfo{author}{\bibfnamefont{B.}~\bibnamefont{Misra}} \bibnamefont{and}
  \bibinfo{author}{\bibfnamefont{E.~C.~G.} \bibnamefont{Sudarshan}},
  \bibinfo{journal}{J. Math. Phys.} \textbf{\bibinfo{volume}{18}},
  \bibinfo{pages}{756} (\bibinfo{year}{1977}).

\bibitem[{\citenamefont{Itano et~al.}(1990)\citenamefont{Itano, Heinzen,
  Bollinger, and Wineland}}]{Itano1990}
\bibinfo{author}{\bibfnamefont{W.~M.} \bibnamefont{Itano}},
  \bibinfo{author}{\bibfnamefont{D.~J.} \bibnamefont{Heinzen}},
  \bibinfo{author}{\bibfnamefont{J.~J.} \bibnamefont{Bollinger}},
  \bibnamefont{and} \bibinfo{author}{\bibfnamefont{D.~J.}
  \bibnamefont{Wineland}}, \bibinfo{journal}{Phys. Rev. A}
  \textbf{\bibinfo{volume}{41}}, \bibinfo{pages}{2295} (\bibinfo{year}{1990}).

\bibitem[{\citenamefont{Erez et~al.}(2008)\citenamefont{Erez, Gordon, Nest, and
  Kurizki}}]{Erez2008}
\bibinfo{author}{\bibfnamefont{N.}~\bibnamefont{Erez}},
  \bibinfo{author}{\bibfnamefont{G.}~\bibnamefont{Gordon}},
  \bibinfo{author}{\bibfnamefont{M.}~\bibnamefont{Nest}}, \bibnamefont{and}
  \bibinfo{author}{\bibfnamefont{G.}~\bibnamefont{Kurizki}},
  \bibinfo{journal}{Nature} \textbf{\bibinfo{volume}{452}},
  \bibinfo{pages}{724} (\bibinfo{year}{2008}).

\bibitem[{\citenamefont{Gordon et~al.}(2009)\citenamefont{Gordon, Bensky,
  Gelbwaser-Klimovsky, Rao, Erez, and Kurizki}}]{Gordon2009}
\bibinfo{author}{\bibfnamefont{G.}~\bibnamefont{Gordon}},
  \bibinfo{author}{\bibfnamefont{G.}~\bibnamefont{Bensky}},
  \bibinfo{author}{\bibfnamefont{D.}~\bibnamefont{Gelbwaser-Klimovsky}},
  \bibinfo{author}{\bibfnamefont{D.~D.~B.} \bibnamefont{Rao}},
  \bibinfo{author}{\bibfnamefont{N.}~\bibnamefont{Erez}}, \bibnamefont{and}
  \bibinfo{author}{\bibfnamefont{G.}~\bibnamefont{Kurizki}},
  \bibinfo{journal}{New Journal of Physics} \textbf{\bibinfo{volume}{11}},
  \bibinfo{pages}{123025} (\bibinfo{year}{2009}).

\bibitem[{\citenamefont{Bensky et~al.}(2010)\citenamefont{Bensky, Rao, Gordon,
  Erez, and Kurizki}}]{Bensky2010}
\bibinfo{author}{\bibfnamefont{G.}~\bibnamefont{Bensky}},
  \bibinfo{author}{\bibfnamefont{D.~D.~B.} \bibnamefont{Rao}},
  \bibinfo{author}{\bibfnamefont{G.}~\bibnamefont{Gordon}},
  \bibinfo{author}{\bibfnamefont{N.}~\bibnamefont{Erez}}, \bibnamefont{and}
  \bibinfo{author}{\bibfnamefont{G.}~\bibnamefont{Kurizki}},
  \bibinfo{journal}{Physica E} \textbf{\bibinfo{volume}{42}},
  \bibinfo{pages}{477} (\bibinfo{year}{2010}).

\bibitem[{\citenamefont{Gordon et~al.}(2010)\citenamefont{Gordon, Rao, and
  Kurizki}}]{Gordon2010}
\bibinfo{author}{\bibfnamefont{G.}~\bibnamefont{Gordon}},
  \bibinfo{author}{\bibfnamefont{D.~D.~B.} \bibnamefont{Rao}},
  \bibnamefont{and} \bibinfo{author}{\bibfnamefont{G.}~\bibnamefont{Kurizki}},
  \bibinfo{journal}{New Journal of Physics} \textbf{\bibinfo{volume}{12}},
  \bibinfo{pages}{053033} (\bibinfo{year}{2010}).

\bibitem[{\citenamefont{Jahnke and Mahler}(2010)}]{Jahnke2010}
\bibinfo{author}{\bibfnamefont{T.}~\bibnamefont{Jahnke}} \bibnamefont{and}
  \bibinfo{author}{\bibfnamefont{G.}~\bibnamefont{Mahler}},
  \bibinfo{journal}{Europhys. Lett.} \textbf{\bibinfo{volume}{90}}, \bibinfo{pages}{50008}
  (\bibinfo{year}{2010}).

\bibitem[{\citenamefont{Breuer et~al.}(2006)\citenamefont{Breuer, Gemmer, and
  Michel}}]{Breuer2006}
\bibinfo{author}{\bibfnamefont{H.-P.} \bibnamefont{Breuer}},
  \bibinfo{author}{\bibfnamefont{J.}~\bibnamefont{Gemmer}}, \bibnamefont{and}
  \bibinfo{author}{\bibfnamefont{M.}~\bibnamefont{Michel}},
  \bibinfo{journal}{Phys. Rev. E} \textbf{\bibinfo{volume}{73}},
  \bibinfo{pages}{016139} (\bibinfo{year}{2006}).

\bibitem[{\citenamefont{Gemmer and Michel}(2006)}]{Gemmer2006}
\bibinfo{author}{\bibfnamefont{J.}~\bibnamefont{Gemmer}} \bibnamefont{and}
  \bibinfo{author}{\bibfnamefont{M.}~\bibnamefont{Michel}},
  \bibinfo{journal}{Eur. Phys. J. B} \textbf{\bibinfo{volume}{53}},
  \bibinfo{pages}{517} (\bibinfo{year}{2006}).

\bibitem[{\citenamefont{Granzow and Mahler}(1998)}]{Granzow1998}
\bibinfo{author}{\bibfnamefont{C.~M.} \bibnamefont{Granzow}} \bibnamefont{and}
  \bibinfo{author}{\bibfnamefont{G.}~\bibnamefont{Mahler}},
  \bibinfo{journal}{Appl. Phys. B} \textbf{\bibinfo{volume}{67}},
  \bibinfo{pages}{733} (\bibinfo{year}{1998}).

\bibitem[{\citenamefont{Landsberg}(1959)}]{Landsberg1959}
\bibinfo{author}{\bibfnamefont{P.~T.} \bibnamefont{Landsberg}},
  \bibinfo{journal}{Physical Review} \textbf{\bibinfo{volume}{115}},
  \bibinfo{pages}{518} (\bibinfo{year}{1959}).

\bibitem[{Not()}]{Note1}
\bibinfo{note}{The potential generalization to larger systems poses a
  challenging question, which should be addressed in future work. We expect
  that, in general, periodic measurements of a thermalizing environment will
  only lead to the undisturbed thermal state of the system, if the repetition
  of measurements is sufficiently slow, such that the effect of the periodic
  reset of the interaction energy is negligible.}

\bibitem[{\citenamefont{\'Alvarez et~al.}(2010)\citenamefont{\'Alvarez, Rao,
  Frydman, and Kurizki}}]{Alvarez2010}
\bibinfo{author}{\bibfnamefont{G.~A.} \bibnamefont{\'Alvarez}},
  \bibinfo{author}{\bibfnamefont{D.~D.~B.} \bibnamefont{Rao}},
  \bibinfo{author}{\bibfnamefont{L.}~\bibnamefont{Frydman}}, \bibnamefont{and}
  \bibinfo{author}{\bibfnamefont{G.}~\bibnamefont{Kurizki}},
  \bibinfo{journal}{Phys. Rev. Lett.} \textbf{\bibinfo{volume}{105}},
  \bibinfo{pages}{160401} (\bibinfo{year}{2010}).

\bibitem[{\citenamefont{Anderson and Qiu}(1997)}]{Anderson1997}
\bibinfo{author}{\bibfnamefont{G.~D.} \bibnamefont{Anderson}} \bibnamefont{and}
  \bibinfo{author}{\bibfnamefont{S.-L.} \bibnamefont{Qiu}},
  \bibinfo{journal}{Proc. Amer. Math. Soc.} \textbf{\bibinfo{volume}{125}},
  \bibinfo{pages}{3355} (\bibinfo{year}{1997}).

\end{thebibliography}
\end{document}